\documentstyle[amssymb,12pt,aps,epsf]{revtex}

\begin{document}
\title{\vspace{2cm} {\large {\bf A Novel Longitudinal Mode in the Coupled Quantum
Chain Compound KCuF$_{3}$. }} \vspace{0.8cm} }
\author{B.\ Lake, D.\ A.\ Tennant and S.\ E.\ Nagler}
\address{Solid State Division, Oak Ridge National Laboratory,\\
Oak Ridge, Tennessee 37831-6393, USA.\\
}
\date{\today}
\maketitle

\begin{abstract}
Inelastic neutron scattering measurements are reported that show a new
longitudinal mode in the antiferromagnetically ordered phase of the spin-1/2
quasi-one-dimensional antiferromagnet KCuF$_{3}$. This mode signals the
cross-over from one-dimensional to three-dimensional behavior and indicates
a reduction in the ordered spin moment of a spin-1/2 antiferromagnet. The
measurements are compared with recent quantum field theory results and are
found to be in excellent agreement. A feature of the data not predicted by
theory is a damping of the mode by decay processes to the transverse
spin-wave branches.
\end{abstract}

\pacs{PACS numbers: 75.10.Jm, 75.10.Hk, 75.30.Ds}


\newpage

Developing~a~comprehensive~understanding~of nonlinear many-body quantum
phenomena is a major objective of condensed matter physics. Low-dimensional
spin systems remain at the forefront of research because large zero-point
fluctuations and non-linearity inherent in the spin commutation relations
create a wealth of exotic quantum phases with unusual dynamics. A model of
central importance is the spin-1/2 ($S$=1/2) Heisenberg antiferromagnetic
chain (HAFC)\ defined by the simple Hamiltonian 
\begin{equation}
{\cal H}_{1D}=J\sum_{i}\vec{S}_{i}\cdot \vec{S}_{i+1}
\end{equation}
where $i$ is the site index and $J$ is the antiferromagnetic exchange
constant. The ground state of the HAFC is a spin singlet, and for
half-odd-integer values of $S$ the natural excitations are free spinons with
a spin of $S$=1/2, not $S$=1 spin-waves as in conventional magnets. Spinons
are restricted to creation in pairs and obey fractional statistics that are
neither Bose-Einstein nor Fermi-Dirac \cite{faddeev81,haldane83}. The spinon
picture has been confirmed in some detail by measurements of the triplet
excitation continuum in KCuF$_{3}$\cite{tennant}, and evidence has also been
seen in a number of other materials\cite{mess}. Thus the dynamics of an
isolated HAFC are relatively well understood theoretically and
experimentally.

Nevertheless, a crucial gap remains in our understanding of real systems
containing embedded HAFCs with inter-chain interactions leading to three
dimensional (3D) antiferromagnetic (AF) order. The long wavelength
excitations are Goldstone modes, so near the antiferromagnetic zone center
(AFZC) one expects well-defined transverse spin-waves obeying a linear
dispersion relation. At energies high compared to the ordering temperature
one-dimensional (1D) quantum fluctuations persist. Exactly how the classical
dynamics characteristic of the ordered state evolve into the high energy
quantum fluctuations is an open question. In light of this, the proposal 
\cite{schulz96,essler97} that a novel longitudinal mode accompanies the
crossover from 1D to 3D physics upon ordering in $S$=1/2 quasi-1D compounds
is of great importance. In this letter we present a neutron scattering study
of such a crossover in KCuF$_{3}$ and confirm the existence of this mode.

The~magnetic~properties~of~KCuF$_{3}$~are~well characterized: It has a
nearly tetragonal crystal structure with lattice parameters $a$=$b$%
=4.126~\AA , and $c$=3.914~\AA (at $T$=10 K). The magnetic Cu$^{++}$ ions
have $S$=1/2 and these are coupled by a strong antiferromagnetic
superexchange interaction ($J\approx 53.5\times 2/\pi $ meV \cite{satija80})
in the $c$-direction. In contrast superexchange in the $a$ and $b$%
-directions is weakly ferromagnetic, $J^{\prime }\approx -10^{-2}J$. The
inter-chain interactions induce magnetic ordering at $T_{N}=39$ K. The spins
are confined to the $ab$ plane, with antiferromagnetic alignment along the
chains ($c$ direction) and ferromagnetic alignment between chains. The
ordered moment-per-spin $m_{0}=\left| <S^{z}>\right| $ (where $z$ is the
ordering direction) is measured to be about $0.27$ \cite{hutchings69} for
temperatures $T\ll T_{N}$, indicating considerable reduction from the
saturation value of 1/2 by zero-point fluctuations.

The one-dimensional dynamics of KCuF$_{3}$ have been studied extensively.
Neutron scattering measurements, made above T$_{N}$ where 1D effects
dominate, showed that the energy ($\omega $) and wavevector ($%
\mbox{\boldmath Q}=[q_{a},q_{b},q_{c}])$ dependence of the spin correlation
functions are in good agreement with spinon model\cite{tennant} as expressed
by the ansatz proposed by M\"{u}ller {\it et.\ al.} \cite{muller81}. Figure
1(a) illustrates the dynamical correlation functions of the M\"{u}ller
ansatz for the energies of interest around the AFZC located at [0,0,-3/2] in
KCuF$_3$; scattering is expected within a V-shaped region centered at $q_c$%
=-3/2. In the three-dimensionally ordered phase below $T_{N}$ the spin
correlations at energies above $\sim$27 meV were not noticibly changed.
However, below this energy well-defined spin-wave modes were found but with
additional scattering lying between them for energies $\omega \gtrsim 12$
meV. Conventional 3D spin-wave theory (SWT) with the inclusion of two-magnon
terms could qualitatively explain the observations at low-energies \cite
{tennant95} but not the continuum at higher energies. Figure 1(b) shows the
scattering calculated from SWT \cite{tennant95} near [0,0,-3/2]. The magnon
branches are well-defined transverse modes, whereas the two-magnon signal is
longitudinal and forms a broad continuum with a maximum at 23.5 meV and a
full-width-half-maximum (FWHM) of 24.0 meV. We note that at this wavevector
unpolarized neutron scattering measures half of the transverse cross-section
plus the full longitudinal cross-section.

Recently, the $T$=0 dynamics of coupled $S$=1/2 HAFCs have been approached
theoretically by considering the solution of the isolated chain in the
continuum limit, treating the inter-chain interactions as a staggered field
and applying the random phase approximation (RPA)\cite{schulz96,essler97}.
The predicted spectrum near the AFZC consists of a doubly degenerate,
well-defined, gapless transverse spin-wave mode, and a well-defined
longitudinal mode with an energy gap, $\Delta _{L}$, proportional to $%
m_{0}^{2}$. The longitudinal mode contributes to the dynamics only when the
ordered moment is suppressed by zero-point fluctuations. Figure 1(c) shows
these theoretical predictions: The longitudinal mode lies between the
dispersions of the transverse modes and has a gap of $\sim $17 meV at the
AFZC. In addition Essler {\it et.\ al.} predict continuum scattering
starting at 22 meV and extending upwards \cite{essler97}. The scan coverage
and resolution in previous measurements \cite{tennant95} was insufficient to
differentiate between the RPA and SWT so new experiments probing the
low-energy sector were necessary to determine the existence of the
longitudinal mode.

Our measurements were performed on the HB1 and HB3 triple-axis spectrometers
located at the High Flux Isotope Reactor, Oak Ridge National Laboratory.
PG(0,0,2) monochromator and analyzer crystals were used and the final energy
was fixed at 13.5 meV with a PG filter placed after the sample to remove
higher order contamination from the beam. Our sample of KCuF$_{3}$ was a
high quality single crystal with mosaic $10^{\prime }$, mass 6.86 g and
volume 1 cm$^{3}$. It was mounted in a variable flow cryostat which provided
a base temperature of 2 K and temperature control to within $\pm 0.1$ K. The
crystal was aligned with the ${\bf a}^{\star }$ and ${\bf c}^{\star }$
reciprocal lattice vectors in the scattering plane and most measurements
took place around the [0,0,-3/2] AFZC. This point was found to be the best
compromise for freedom from phonon background, and maximization of
longitudinal magnetic intensity. With the collimation 48$^{\prime }$-40$%
^{\prime }$-40$^{\prime }$-240$^{\prime }$ the transverse modes have a FWHM
of 0.057 \AA $^{-1}$ (0.03525 r.l.u.) and the energy resolution was 1.3 meV
FWHM at 16 meV energy transfer.

Several constant-energy and constant-wavevector scans were made over the
energies 7 to 27 meV and the wavevectors $l$=-1.65 to -1.35 r.l.u. These
scans were measured at a number of temperatures ranging from well below $%
T_{N}$ at 2 K to well above it at 200 K. Figure 2(a) shows two
constant-wavevector scans made at the AFZC [0,0,-3/2]. The filled circles
are data measured at $T\ll T_{N}$ in the temperature range 2 K to 10 K. To
distinguish magnetic from nonmagnetic features, further measurements were
made at 200 K (open circles). At such high temperatures magnetic scattering
is washed out by thermal broadening identifying the peaks at 19.5 meV and 25
meV as phonons.

To display the magnetic contributions at $T\ll T_{N}$ more clearly, we show
the $T\leq $11 K data in Fig.\ 2(b) with the non-magnetic contributions
subtracted off (the 200 K lineshape was subtracted but with the phonon
amplitudes adjusted for the thermal population factor). Two features
dominate the scattering:\ a large signal at low energies and a broad peak
centered near 16 meV. The low-energy scattering comes from the capture of
the transverse modes by the resolution function; while the broad peak lies
at the anticipated longitudinal mode energy. The solid line is the fit of
the instrumental resolution convolved with the longitudinal mode dispersion 
\cite{essler97} assuming a Gaussian profile and a $1/\omega $ structure
factor. The fitted parameters were $\Delta _{L}$, the peak width and an
overall intensity. The fitted $\Delta _{L}$ was 14.9$\pm $0.1 meV which, due
to resolution effects, is lower than the apparent mode position of $\sim $16
meV. This gap energy is similar to the theoretical value of 17.4 meV \cite
{essler97} but quite different from the position of the two-magnon maximum
calculated from SWT which should appear at 23.5 meV. The mode is
intrinsically broadened with a FWHM of 4.95$\pm $0.07 meV suggesting that
the lifetime may be shortened by decaying into spin-waves. It should be
noted that this feature is still much sharper than two-magnon scattering
predicted by SWT.

The longitudinal mode associated with ordering in the quasi-1D, spin-1/2
HAFC should not exist above the transition temperature. The scan made at 200
K (Fig.\ 2(a)) shows that the observed mode is not present at temperatures $%
T\gg T_{N}$. This measurement was repeated close to $T_{N}$ over the range
30 to 40 K and is displayed in Fig.\ 2(c). It shows increased scattering at
low energies compared to the $T\ll T_{N}$ data, with the region between the
transverse and longitudinal modes filled in. The longitudinal mode is
indistinguishable from the smooth continuum scattering. Previous experiments 
\cite{tennant} have shown that the scattering above $T_{N}$ is consistent
with that expected for an ideal 1D chain. As the AF order is reduced one
expects the free spinon continuum scattering to fill in at lower energies.

It is important to eliminate the possibility that the feature at 16 meV is
signal from the transverse branches that has been distorted by the
resolution function to give the appearance of a mode. Fig.\ 3 shows constant
energy scans at 10 meV (a) and 16 meV (b). Each scan shows two peaks which
come from the transverse branches, and the solid lines are fits of the
transverse mode dispersion (Fig.\ 1(c)) convolved with the instrumental
resolution, where the only fitted parameter is the amplitude. At 10 meV the
fit is remarkably good and shows the accuracy with which the resolution is
known. At 16 meV, where the longitudinal mode is seen in the
constant-wavevector scan of Fig.\ 2(b), the profile can no longer be fitted
by the transverse modes alone. The extra scattering occurring in between the
peaks demonstrates that the feature at 16 meV cannot originate from the
transverse modes. Additional scans at other wavevectors where the resolution
is different were also used to confirm that the 16 meV scattering was not an
artifact.

Finally a series of constant-wavevector scans were performed at 10 K to map
out the scattering as a function of energy and wavevector over the ranges 8
to 22.5 meV and -1.65 to -1.35 r.l.u. The phonon at 19.5 meV (Fig.\ 2(a))
was modelled and subtracted using additional scans at 200 K and the
resulting data is displayed as a contour plot in Fig.\ 4(a) The transverse
modes form the red V-shaped rods dispersing from [0,0,-3/2] and the
longitudinal mode is the yellow band lying between the transverse branches.
Figure 4(b) shows a simulation of the predicted magnetic scattering over the
same energy and wavevector region. The longitudinal mode was assumed to
follow the theoretical dispersion \cite{essler97}, except that the
zone-centre energy gap was fixed at 14.9 meV as obtained from fitting Fig.\
2(b). The lineshape of the mode was Gaussian with a FWHM of 4.95 meV (also
extracted from the fit), while the transverse modes were given a resolution
limited profile. Care was taken to ensure that the lineshapes of the modes
were properly normalized so the that the integrated intensity of the
tranverse modes was four times greater than that of the longitudinal mode as
predicted theoretically \cite{essler97}. The calculation includes the
thermal correction, the neutron polarization term and the Cu$^{++}$ magnetic
form factor. The resemblance between Fig.\ 4(a) and Fig.\ 4(b) is striking
and demonstrates not only the very real presence of the longitudinal mode,
but also the accuracy with which the theories predict its intensity relative
to the transverse modes.

In summary our experiments have established the presence of a novel mode in
KCuF$_{3}$, with an energy and intensity quantitatively consistent with the
predictions of RPA theory for a longitudinal mode in coupled S=1/2 chains.
The mode was found to have a broadened linewidth suggesting an instability
to decay to spin-waves. Longitudinal modes should occur in other quasi-1D
S=1/2 spin systems that show significant zero-point reduction in their
ordered moment. Interestingly, a similar longitudinal mode is also predicted
when the spectrum of the S=1/2 HAFC in a staggered field is calculated using
pseudo-boson dimer operators\cite{lake97}. The dimer basis approach
preserves most of the quantum fluctuations that are discarded by SWT and
underscores that the physical origin of the longitudinal mode is the zero
point fluctuations that suppress the ordered moment in the coupled chain
system. Finally, we note that an isolated longitudinal mode is also present
in the Haldane-gapped (S=1) chain in the presence of a staggered field\cite
{ray99}, but the physics there is somewhat different since this corresponds
to a splitting of the well defined gap mode.

We are grateful to G.\ Shirane for the loan of the sample, and to A.M.\
Tsvelik for helpful discussion. D.A.T.\ would like to thank Ris\o\ National
Laboratory for financial support and hospitality as a visiting scientist.
ORNL is managed for the U.S. D.O.E. by Lockheed Martin Energy Research under
contract no. DE-AC05-96OR22464.

\begin{figure}[tbp]
\caption{ Three theories of the magnetic correlations in KCuF$_{3}$ plotted
as functions of wavevector and energy. (a) shows the two-spinon continuum
given by the M\"{u}ller ansatz \protect\cite{muller81}, where the intensity
of the scattering is indicated by the shading of the contours. (b) shows the
scattering from SWT which gives both transverse modes (thick black lines)
and a two-magnon continuum (intensity indicated by the shading) \protect\cite
{tennant95}. (c) shows the longitudinal mode predicted by field theory 
\protect\cite{essler97}. }
\end{figure}

\begin{figure}[tbp]
\caption{ Constant-wavevector scans at [0,0,-3/2]. (a) The scattering at $%
T<11$ K (closed circles) compared to the scattering at 200 K (open circles),
both magnetic and phonon signal is observed. The lines are guides to the
eye. (b) The magnetic scattering at $T<11$ K with the phonon background
subtracted off. The solid line is a fit as described in the text and the
dashed line is a guide to the eye showing the tail of the transverse
scattering. (c) The magnetic scattering measured in the temperature range
30-40 K, again the phonons have been subtracted. The line is a guide to the
eye. }
\end{figure}

\begin{figure}[tbp]
\caption{ Constant-energy scans measured at 10 K for the energies (a) 10 meV
and (b) 16 meV. The solid line is a fit of the transverse modes convolved
with the resolution function. }
\end{figure}

\begin{figure}[tbp]
\caption{ (color) (a) shows an energy-wavevector contour map of the magnetic
signal collected at 10 K; the colors indicate the relative scattering
intensities. (b) shows a simulation of the magnetic signal over the same
reciprocal space region using the theoretical dispersions for the transverse
and longitudinal modes \protect\cite{essler97} convolved with the resolution
function. }
\end{figure}


\begin{references}
\bibitem{faddeev81}  L.D. Faddeev and L.A. Takhtajan, Phys. Lett. {\bf 85A},
375 (1981).

\bibitem{haldane83}  F.D.M. Haldane and M.R. Zirnbauer, Phys. Rev. Lett. 
{\bf 71}, 4055 (1993).

\bibitem{tennant}  S.E. Nagler {\it et. al.} Phys. Rev. B {\bf 44}, 12361
(1991), D.A. Tennant, {\it et. al.} Phys. Rev. Lett {\bf 70}, 4003 (1993),
D.A. Tennant, {\it et. al.} Phys. Rev. B {\bf 52}, 13368 (1995);

\bibitem{mess}  I.U. Heilman {\it et. al.}, Phys. Rev. B. {\bf 18}, 3530
(1978), D.C. Dender {\it et. al.}, Phys. Rev. Lett{\em .} {\bf 79}, 1750
(1997), R. Coldea {\it et. al.}, Phys. Rev. Lett. {\bf 79}, 151 (1997), I.
Tsukuda {\it et. al.}, Phys. Rev. B. {\bf 60}, 6601 (1999).

\bibitem{schulz96}  H.J. Schulz,Phys. Rev. Lett. {\bf 77}, 2790 (1996).

\bibitem{essler97}  F.H.L. Essler, A.M. Tsvelik, and G. Delfino, Phys. Rev.
B. {\bf 56}, 11001 (1997).

\bibitem{satija80}  S.K. Satija {\it et. al.}, Phys. Rev. B {\bf 21}, 2001
(1980).

\bibitem{hutchings69}  M.T. Hutchings {\it et. al.}, Phys. Rev. {\bf 188},
919 (1969).

\bibitem{muller81}  G. M\"{u}ller, {\it et. al.},Phys. Rev. B {\bf 24} 1429
(1981).

\bibitem{tennant95}  D.A. Tennant, {\it et. al.} Phys. Rev. B {\bf 52},
13381 (1995).

\bibitem{lake97}  B. Lake, R.A. Cowley, and D.A. Tennant, J. Phys.: Condens.
Matter {\bf 9}, 10951 (1997).

\bibitem{ray99}  S. Raymond {\it et. al.}, Phys. Rev. Lett. {\bf 82}, 2382
(1999).
\end{references}
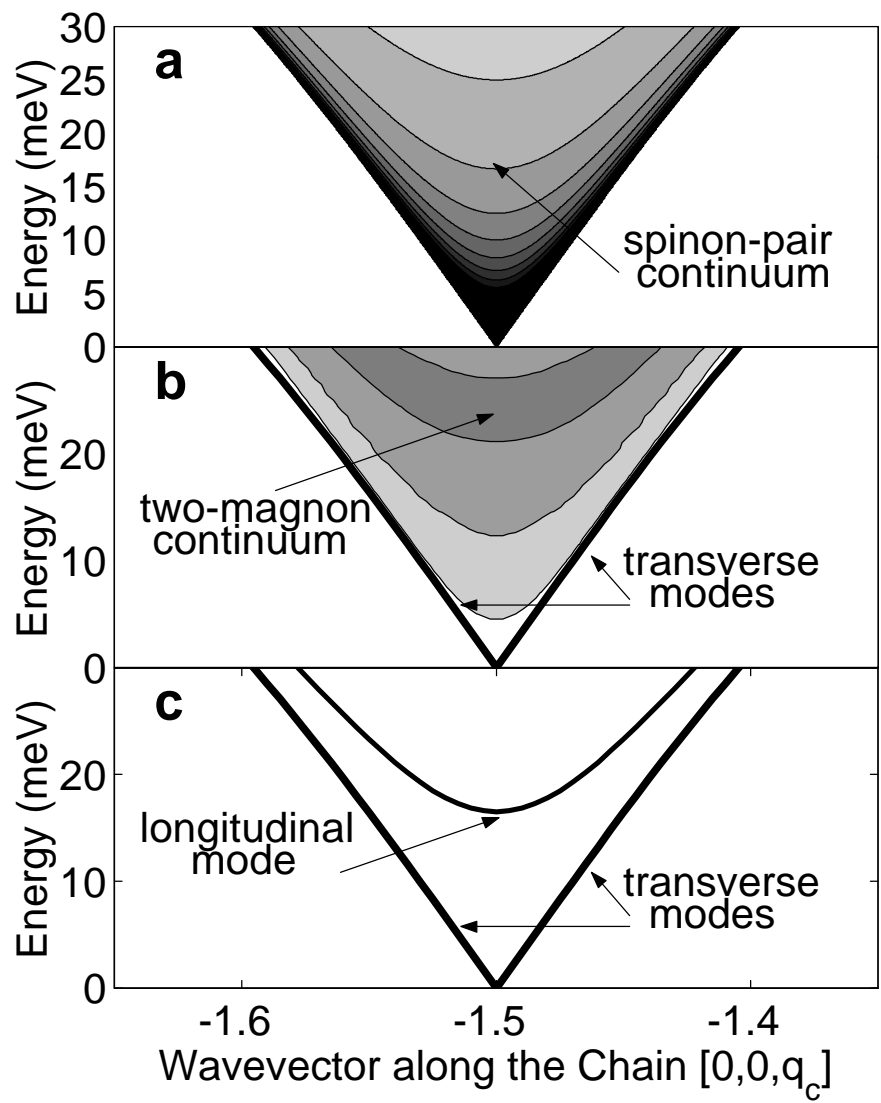
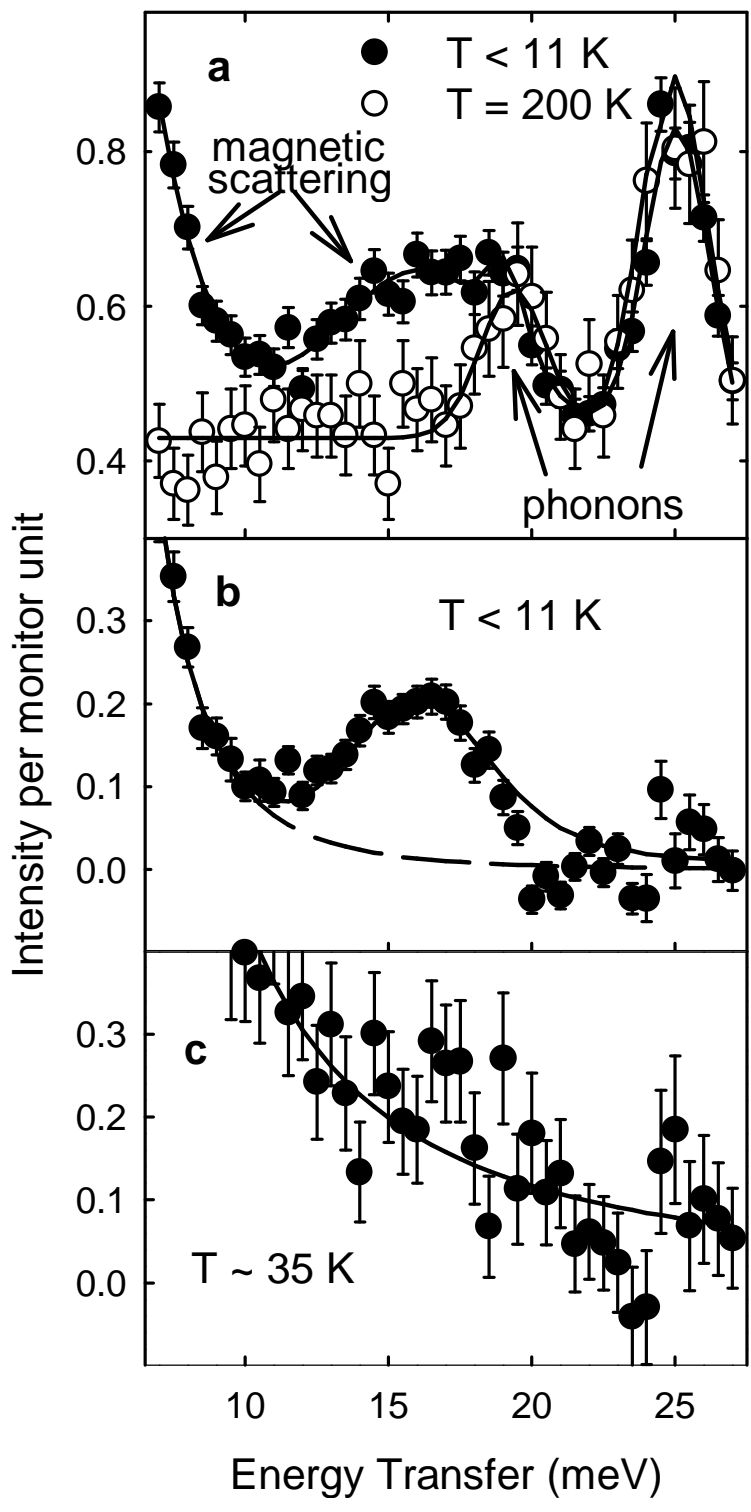
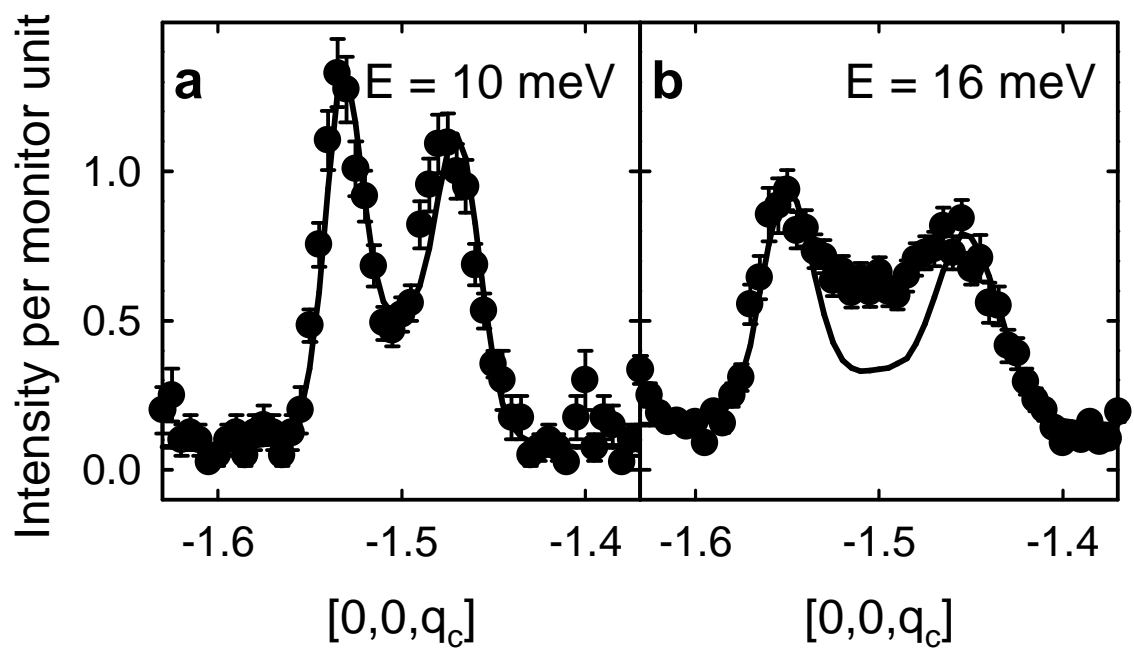
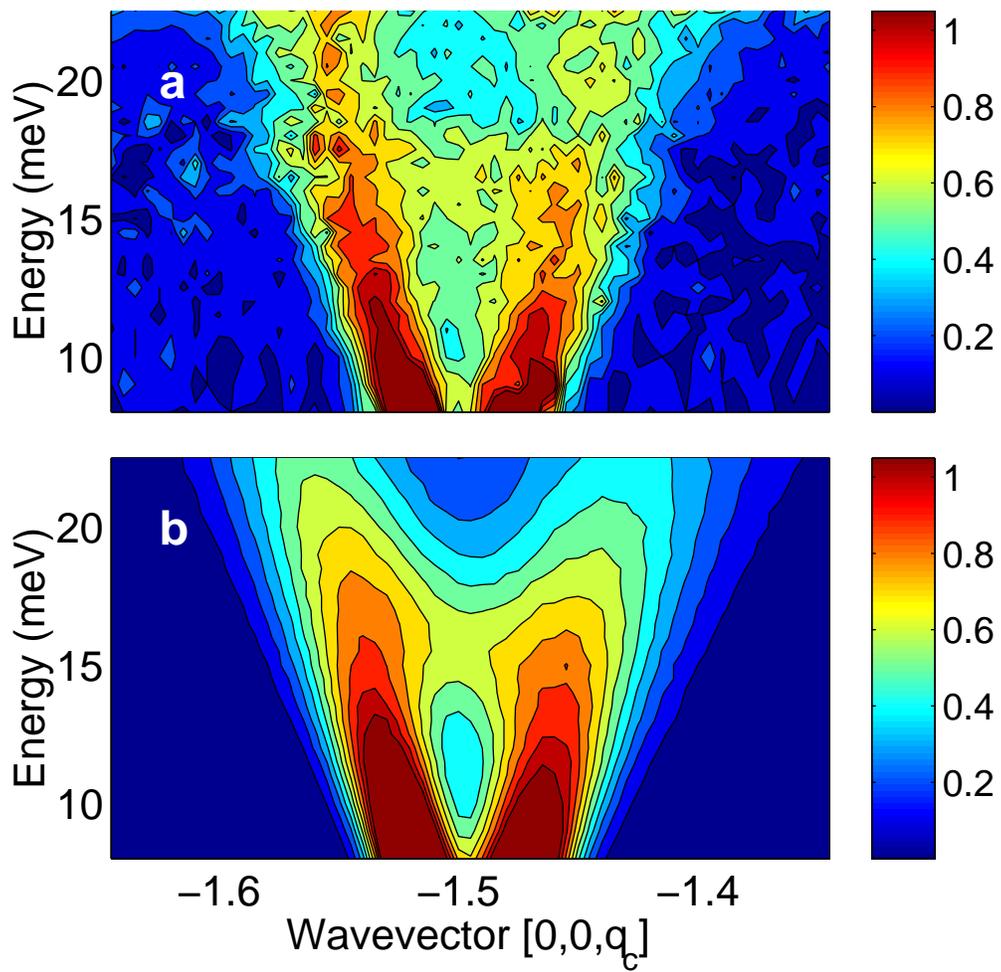
\end{document}